\documentclass[preprint,authoryear,12pt]{elsarticle}

\usepackage{graphicx}
\usepackage{stfloats}\usepackage{subfigure}
\usepackage{caption}

\usepackage{amssymb}
\usepackage{amsthm}
\usepackage{amsmath} 

\usepackage{verbatim}
\usepackage{cite}
\usepackage{epstopdf}
\usepackage{acronym}
\usepackage{ifthen}

\begin{document}


\newtheorem*{quest}{Question}
\newtheorem{df}{Definition}
\newtheorem{eg}{Example}
\newtheorem{cor}{Corollary}
\newtheorem{thm}{Theorem}
\newtheorem{conj}{Conjecture}
\newtheorem{lem}{Lemma}

\newenvironment{ans}{\par\indent \emph{Ans:}}{\noindent \qedsymbol}    

\acrodef{VA}{Variable Annuity}
\acrodef{MC}{Monte Carlo}
\acrodef{IDW} {Inverse Distance Weighting}
\acrodef {RBF} {Radial Basis Function}
\acrodef {AV} {Account Value} 
\acrodef {GD} {Guaranteed Deat Benefit Value} 
\acrodef {CDF} {Cumulative Distribution Function}
\acrodef {LHS} {Latin Hypercube Sampling}
\acrodef {LSMC} {Least Squares Monte Carlo}
\acrodef {SCR} {Solvency Capital Requirement}

\def\bibsection{\section*{References}}

\begin{frontmatter}



\title{{\bf A Spatial Interpolation Framework for Efficient Valuation of Large Portfolios of Variable Annuities}}


\author[CSUT]{Seyed Amir Hejazi}
\ead{amir@cs.toronto.edu}
\author[CSUT]{Kenneth R. Jackson}
\ead{krj@cs.toronto.edu}
\address[CSUT]{Department of Computer Science, University of Toronto, Toronto, ON, M5S 3G4, Canada}
\author[MUCONN]{Guojun Gan}
\ead{Guojun.Gan@uconn.edu}
\address[MUCONN]{Department of Mathematics, University of Connecticut, Storrs, Connecticut, 06269-3009, USA}

\begin{abstract}
\ac{VA} products expose insurance companies to considerable risk because of 
the guarantees they provide to buyers of these products. Managing and hedging these risks requires 
insurers to find the value of key risk metrics for a large portfolio of \ac{VA} products. 
In practice, many companies rely on nested \ac{MC} simulations to find key risk metrics. 
\ac{MC} simulations are computationally demanding, forcing insurance companies to invest 
hundreds of thousands of dollars in computational infrastructure per year. Moreover, 
existing academic methodologies are focused on fair valuation of a single \ac{VA} 
contract, exploiting ideas in option theory and regression. 
In most cases, the computational complexity of these methods surpasses 
the computational requirements of \ac{MC} simulations.
Therefore, academic methodologies cannot scale well to large portfolios of \ac{VA} contracts. 
In this paper, we present a framework for valuing such portfolios based on spatial interpolation. We provide 
a comprehensive study of this framework and compare existing interpolation schemes. Our numerical results 
show superior performance, in terms of both computational efficiency and accuracy, for these methods compared 
to nested \ac{MC} simulations. We also present insights into the challenge of finding an effective 
interpolation scheme in this framework, and suggest guidelines that help us build a fully automated scheme 
that is efficient and accurate. 

\end{abstract}


\end{frontmatter}

\section{Introduction}\label{sec:intro}

Variable annuities are unit-linked products that are wrapped with a 
life insurance contract. These products allow a policyholder to invest into 
pre-defined sub-accounts set up by the insurance company. Sub-account funds 
are invested in bonds, the money market, stocks and other financial products. An insurer
offers different types of sub-accounts that are tailored to the appetite of policyholders 
with different level of tolerance for risk. The investment can be made via 
a lump-sum payment or a series of investment purchases. In return, the insurance company offers tax 
sheltered growth and  guarantees that protect the policyholder in a bear market \citep{Chi12}.

Upon entering into a contract, the policyholder is given two accounts: one keeps track of the 
performance of investments in the sub-accounts and the other keeps track of the amount of guarantee provided 
by the insurance company. The value of former account is called the account value and the value of 
latter account is called the benefit base. 
During a period called the accumulation phase, the policyholder accumulates assets on his investments 
in sub-accounts and the value of his benefit base appreciates by contractually agreed 
roll ups, ratchets and resets without taxation. At the 
end of the accumulation phase, the benefit base is locked in and the insurer guarantees to 
return at least the benefit base as a lump sum payment 
or as a stream of payments during a period called the withdrawal phase.

The most prevalent of the guarantees are the Guaranteed Minimum Death Benefit (GMDB), 
the Guaranteed Minimum Withdrawal Benefit (GMWB), the Guaranteed Minimum Income Benefit (GMIB), 
and the Guaranteed Minimum Accumulation Benefit (GMAB). 
The GMDB guarantees a specified lump sum payment on death regardless of the performance of 
the underlying account. The most basic guarantee offered now is the return of 
benefit base adjusted for any partial withdrawals. The GMWB  guarantees the 
ability to partially withdraw up to a pre-determined percentage (called the withdrawal rate) 
of the benefit base for a specified number of years. The decision to withdraw is made 
annually and the maximum amount of withdrawal is a function of the age of the policyholder. 
The GMIB guarantees a stream of income for life contingent on the survival of the policyholder, and the GMAB 
guarantees a lump sum payment on maturity of the contract regardless of the performance of the 
underlying funds. For further details, see \citep{Geneva13}. 

Embedded guarantees are the key selling feature of \ac{VA} products. These 
guarantees have allowed insurance companies to sell trillions of dollars worth of these 
products worldwide, in 2010 alone \citep{IRI11}. As a result, major insurance 
companies are now managing large portfolios of \ac{VA} contracts, each with  
hundreds of thousands of contracts.

Although the embedded guarantees are attractive features to the buyer of \ac{VA} products, they 
expose the insurers to substantial risk (e.g., market and behavioral risk). 
Because of that, major insurance companies have started risk management and hedging \citep{Boyle97, Hardy03}
programs, especially after the market crash of 2008, to reduce their exposures. 
An integral part of a risk management program is finding the value of key statistical 
risk indicators, e.g., the Greeks \citep{Hull06}, and the \ac{SCR} \citep{Bauer12}, on daily, monthly 
and quarterly bases. 

Most of the academic research to date has focused on fair valuation of individual 
\ac{VA} contracts \citep{Coleman06, Milevsky06, Boyle08, Lin09, Belanger09, Gerber12, 
Chen08, Dai08, DHalluin05, Azimzadeh13, Chen08-2, Huang11, Moenig11, Ulm06}. 
Most of the methodologies developed in these research papers are based on 
ideas from option pricing theory, and are tailored to the type of studied \ac{VA}. 
In addition, almost all of the proposed schemes are computationally expensive and 
the results they provide for a \ac{VA} contract cannot be re-used for another 
\ac{VA} contract, even of similar type. 
Each \ac{VA} contract is unique in terms of its key attributes, i.e., 
age, gender, account value, guaranteed value, maturity of contract, 
fund type, etc. Hence, \ac{VA} portfolios are non-homogeneous pools of \ac{VA} contracts, and 
academic methodologies cannot scale well to be used to calculate key risk statistics of 
large \ac{VA} portfolios.

Although the nature of the guarantees in the \ac{VA} products makes them path dependent, in practice, 
insurance companies relax the assumptions on guarantees and rely heavily on stochastic 
simulations to value these products and manage their risks. In particular, nested \ac{MC} 
simulations are the industry standard methodology in determining key risk metrics \citep{Bauer12, Reynolds08}. 
Nested simulations consist of outer loops that span the space of key market variables (risk factors), and 
inner loops that project the liability of each \ac{VA} contract along many simulated risk-neutral paths \citep{Fox13}. 
As explained in Section \ref{sec:ne}, \ac{MC} simulations are computationally demanding, 
forcing insurance companies to look for ways to reduce the computational 
load of \ac{MC} simulations. 

In this paper, we focus on the issue of efficient approximation of the Greeks for large portfolios of \ac{VA} products. 
In particular, we provide an extensive study of a framework based on metamodeling that can 
approximate the Greeks for large portfolios of \acp{VA} in a fast and accurate way. The rest of this paper 
is organized as follows. Section \ref{sec:literature} reviews existing efforts to reduce the 
computational requirements of \ac{MC} simulations. Section \ref{sec:sif} introduces the spatial 
interpolation framework that offers fast and accurate valuation of large \ac{VA} portfolios. We  
discuss various ways that this method can be employed with existing interpolation schemes to approximate 
the Greeks of \ac{VA} portfolios. Section \ref{sec:ne} provides insights into the performance of 
the proposed framework via extensive simulations. Finally, Section \ref{sec:conclusion} concludes 
the paper.

\section{Review of Existing Methods}\label{sec:literature}
\acp{VA} can be considered to be a type of exotic market instrument \citep{Hull06}. Hence, in 
this section, we provide a summary of the main existing approaches to the 
more general problem of valuing large portfolios of exotic market instruments. 
As we discussed earlier, 
\ac{MC} simulations are the industry standard approach to value path dependent products. The 
problem with \ac{MC} simulations is that they are computationally demanding and do not 
scale well to large portfolios of exotic instruments. The existing methods try to alleviate 
the computational burden of \ac{MC} simulations by approximating the surface that defines 
a mapping between the key risk metrics of interest and the key economic factors. Most of these 
approaches are based on ideas in regression theory and require a time consuming calibration 
to find the regression coefficients. 

A well-studied approach in the literature is the method of replicating portfolios. 
In a replicating portfolio, the main idea is to regress the cash flow of the input portfolio, at 
each time step, against the cash flow of a portfolio of well-formulated financial instruments, 
in particular, vanilla derivatives. The financial instruments that are used in 
regression often have closed form formulas for liability values. This simplifies the 
calculation of cash flows and subsequently reduces the cost of finding 
regression coefficients. The problem of determining a replicating portfolio is often formulated as 
a minimization problem with respect to a norm. Depending on the choice of norm, 
quadratic programming \citep{Daul09, Oechslin07} and linear programming \citep{Dembo99} approaches 
have been proposed to solve the optimization problem. However, 
constructing the replicating portfolio via either method is time consuming because it requires 
projecting the cash flow at each time step. Moreover, depending on the desired level of 
accuracy, the number of regression variables can grow polynomially with the number of attributes of 
the instruments in the input portfolio. Another major issue with replicating portfolios is the 
difficulty in incorporating the actuarial risk factors. The usual practice in the literature is to assume 
that these values follow their expected value. 

Another important regression approach is the method of \ac{LSMC}. 
In \ac{LSMC}, the liability value is regressed on key economic factors \citep{Cathcart09}.
The common practice is to choose powers of key economic factors as basis functions and 
approximate the liability as a linear combination of the basis functions \citep{Carriere96, Longstaff01}. 
Hence, the Greeks of the input portfolio can be calculated 
as the sum of the estimated Greeks for each \ac{VA} in the input portfolio. 
In order to find regression coefficients, \ac{LSMC} uses poor \ac{MC} estimates 
of liability at each point in the space of regression variables. Cathcart and Morrison 
\citep{Cathcart09} provide examples in which the number of 
\ac{MC} projection paths to estimate liability at each point in the space is reduced to one 
and yet the regression function provides fairly accurate estimates. However,  
to achieve a reasonable accuracy, \ac{LSMC} usually requires many sample
points for numeric explanatory variables that can assume values in 
a large interval. Each \ac{VA} product has several numeric attributes, some of which are unique to the 
type of \ac{VA} contract. Therefore, the \ac{LSMC} method incurs significant computational costs 
to accurately regress the liability of contracts in a portfolio consisting of different types of VA products.

Recently, a new spatial functional data analysis approach \citep{Gan15, Gan13-2, Gan13-3} has 
been proposed that addresses the computational complexity of 
nested simulations by reducing the number of \ac{VA} contracts included in 
the \ac{MC} simulations. The proposed methods first select a small set of representative
\ac{VA} policies, using various data clustering \citep{Gan07} or sampling methods, and price only these 
representative policies via \ac{MC} simulations. The representative contracts 
and their Greeks are then fed as training samples to a machine learning algorithm \citep{Bishop06} 
called Kriging \citep{Cressie93}, which then estimates the Greeks of the whole portfolio. 
In the rest of the paper, we provide a study of the more general framework 
of spatial interpolation, including Kriging methods, and provide more 
insights into why spatial interpolation can be much more efficient and accurate than other 
approaches in the literature. In this paper, we use the term interpolation in 
the general context of estimating the values at unknown locations using the known data at 
a finite number of points. In this context, an interpolation method that exactly reproduces the 
interpolated values is called an \emph{exact} interpolator. 
Interpolation methods that do no satisfy this constraint are called \emph{inexact} 
interpolation methods \citep{Burrough98}.

\section{Spatial Interpolation Framework}\label{sec:sif}
The proposed methods in \citep{Gan15, Gan13-2, Gan13-3} can be categorized 
under the general framework of spatial interpolation. Spatial interpolation 
is the procedure of estimating the value of data at unknown locations 
in space given the observations at sampled locations \citep{Burrough98}. As 
the definition suggests, spatial interpolation requires a sampling method 
to collect information about the surface of interest and an interpolation 
method that uses the collected information to estimate the value of the surface at 
unknown locations. As discussed in \citep{Gan15, Gan13-2, Gan13-3}, the choice of 
sampling method and interpolation method can noticeably impact the quality of the 
interpolation. In this paper, we choose to focus on the latter, and leave for another 
paper a discussion of the choice of an appropriate sampling method. 

In the functional data analysis literature, there exist two main classes of interpolation 
methods \citep{Burrough98}:

\begin{itemize}
 \item {\bf Deterministic Interpolation:} Creates surfaces from measured points on the basis of 
 either similarity or degree of smoothness. 
 \item {\bf Stochastic Interpolation:} Utilizes statistical properties of measured points, such as 
 auto-correlation amongst measured points, to create the surface.
\end{itemize}

In what follows, we study three (one stochastic, and two deterministic) of the most prominent 
of these interpolation techniques--- Kriging, \ac{IDW} and \ac{RBF}--- in the context of our problem 
of interest. In particular, we study how well these interpolation techniques 
estimate the delta value for a large portfolio of \ac{VA} products. Although our study focuses 
on estimation of the delta value, the framework is general and 
can be applied to estimate other Greeks as well. We compare the performance 
of these methods in terms of computational complexity and accuracy at the micro (contract) level  and 
at the macro (portfolio) level. 

Although \citep{Gan15, Gan13-2} provide some insights into the performance of the Kriging interpolation 
methods, we provide further insights into the efficiency and accuracy of Kriging based methods in 
comparison to other interpolation techniques. Moreover, we shed some light on how Kriging achieves its 
documented performance and discuss some issues regarding the choice of variogram model and distance function.  

\subsection{Sampling Method}\label{sec:sampling_method}
In this paper, we focus on studying synthetic portfolios that are generated uniformly at random 
in the space of selected variable annuities. 
In \citep{Gan13-3}, the \ac{LHS} method \citep{McKay79} is used to select representative contracts. \ac{LHS} provides a 
uniform coverage of the space including the boundary \ac{VA} contracts.
The results of \citep{Gan13-3} indicate that LHS increases the accuracy of the estimation compared to 
other sampling methods. In order to preserve the properties of LHS, we select our representative contracts 
by dividing the range of each numeric attribute of a \ac{VA} contract into almost equal length subintervals, 
selecting the end points of resulting subintervals and producing synthetic contracts from all combinations of 
these end points and choices of categorical attributes.

\subsection{Kriging}\label{sec:kriging}

Kriging is a stochastic interpolator that gives the best linear 
unbiased estimation of interpolated values  assuming a Gaussian process 
model with prior covariance \citep{Matheron63, Krige51}. 
Various Kriging methods (i.e., Simple Kriging, Ordinary Kriging, Universal 
Kriging, etc.) have been developed based on assumptions about the 
model. In our experiments, we didn't find any 
significant advantages in choosing a particular Kriging method. Therefore, 
for the sake of simplicity of analysis, and based on the results of \citep{Gan15}, 
we choose to study ordinary Kriging in this paper. 

Assume $Z(x)$ represents the delta value of a \ac{VA} contract represented in 
space by the point $x$. Let $Z(x_1), Z(x_2), \ldots, Z(x_n)$ represent the 
observed delta values at locations $x_1, x_2, \ldots, x_n$. Ordinary Kriging tries 
to find the best, in the Mean Squared Error (MSE) sense,  unbiased linear 
estimator $\hat{Z}(x) = \sum_{i = 1}^n \omega_i Z(x_i)$ of $Z(x)$ by 
solving the following system of linear equations to find the $w_i$s.

\begin{equation}
 \label{eq:ord_krig}
 \begin{bmatrix}
  \gamma(D(x_1, x_1)) & \gamma(D(x_1, x_2)) & \ldots & \gamma(D(x_1, x_n)) & 1\\
  \vdots & \vdots & \ddots & \vdots & \vdots\\
  \gamma(D(x_n, x_1)) & \gamma(D(x_n, x_2)) & \ldots & \gamma(D(x_n, x_n)) & 1\\
  1 & 1 & \ldots & 1 & 0
 \end{bmatrix}
 \begin{bmatrix}
  w_1\\ \vdots \\ w_n \\ \lambda
 \end{bmatrix}
 = 
 \begin{bmatrix}
  \gamma(D(x_1, x)) \\ \ldots \\ \gamma(D(x_n, x)) \\ 1
 \end{bmatrix}
\end{equation}
where $\lambda$ is the Lagrange multiplier \citep{Boyd04}, $\gamma(\cdot)$ is the semi-variogram 
function, to be discussed shortly, and $D(\cdot,\cdot)$ is a distance function that measures 
the distance between two points in the space of \ac{VA} contracts. The last row 
enforces the following constraint to allow an unbiased estimation of 
$Z(x)$. 

\begin{equation}
 \sum_{i = 1}^n w_i = 1
\end{equation}

In this formulation of the Kriging problem, the system of linear equations \eqref{eq:ord_krig} 
should be solved once for each \ac{VA} policy (point in space). 
Solving a system of linear equations, with standard methods, takes $\Theta(n^3)$\footnote{$f(x) = \Theta(g(x))$ means that there exists 
positive numbers $c_1$, $c_2$, and $M$ such that $\forall x > M: c_1 g(x) \leq f(x) \leq c_2 g(x)$.} time. 
Hence, estimating the delta value for a portfolio of $N$ \ac{VA} contracts by 
summing the estimated delta value of each contract requires $\Theta(N \times n^3)$ time 
which is computationally expensive. Because of this, Kriging methods are 
inefficient in finding a granular view of the portfolio. However, if we are only 
interested in the Greeks of the portfolio, and not the Greeks of each individual 
policy, we can follow the approach of \citep{Gan15, Gan13-2, Gan13-3} and use the 
linearity of the systems of linear equations to sum them 
across the portfolio in $\Theta(n \times N)$ time and to solve only the resulting system of 
linear equations in time proportional to $n^3$. Hence estimating the delta of a portfolio requires 
$\Theta(n^3 + n \times N)$ time. 
To sum the systems of linear equations, we  
sum the corresponding weights and Lagrange multipliers 
on the left hand side of the equations and sum the corresponding 
terms, i.e., $\gamma(D(x_i, x)), i = 1, 2, \ldots, n$, and constants, on the right hand 
side of the equations.

\subsubsection{Variogram}
Kriging assumes the Gaussian process $Z(x)$ is second order stationary, i.e., the covariance 
of the Gaussian process in two locations is a function of distance between the two locations. 
Assuming a zero mean, the Gaussian process covariance function can be defined in terms of 
a variogram function $2\gamma(h)$:

\begin{align}
 Cov_{Z}(x + h, x) = &E\big[Z(x + h) Z(x)\big] \nonumber \\
 = &\frac{1}{2}E\Big[Z^2(x + h) + Z^2(x) - \Big(Z(x + h) - Z(x)\Big)^2\Big] \nonumber \\
 = &Var(Z) - \frac{1}{2} (2\gamma(h))
\end{align}

In practice, for a set of sample points $x_i, 1\leq i\leq n$, the variogram
can be estimated as 

\begin{equation}\label{eq:empirical_vario}
 2\hat{\gamma}(h) = \frac{1}{N(h)} \sum_{i = 1}^{N(h)} \big(Z(x_i + h) - Z(x_i)\big)^2
\end{equation}
where $N(h)$ is the number of pairs in the sample separated by a distance h from each other. 
The function $2\hat{\gamma}(h)$ is often called the \emph{empirical variogram}.
 
Because of the noise in measurements, the estimated empirical variogram may not represent a valid 
variogram function. Since methods like Kriging require valid variograms at every distance h, 
empirical variograms are often approximated by model functions ensuring the validity of the 
variogram \citep{Chiles99}. 
Variogram models are usually described in terms of three important variables:

\begin{itemize}
 \item {\bf Nugget (n):} The height of the discontinuity jump at the origin. 
 \item {\bf Sill (s):} The Limit of the variogram as the lag distance $h$ approaches infinity. 
 \item {\bf Range (r):} The distance at which the difference of the variogram from the sill becomes negligible. 
\end{itemize}

Figure \ref{fig:variogram} shows an example of an empirical variogram and the model variogram.  
In our study, we choose to focus on the following three prominent variogram models \citep{Chiles99, Cressie93}:

\begin{figure}[!bt]
 \centering
 \includegraphics[trim=1cm 1cm 1cm 1cm, clip=true, width=0.7\textwidth, height=0.4\textheight]{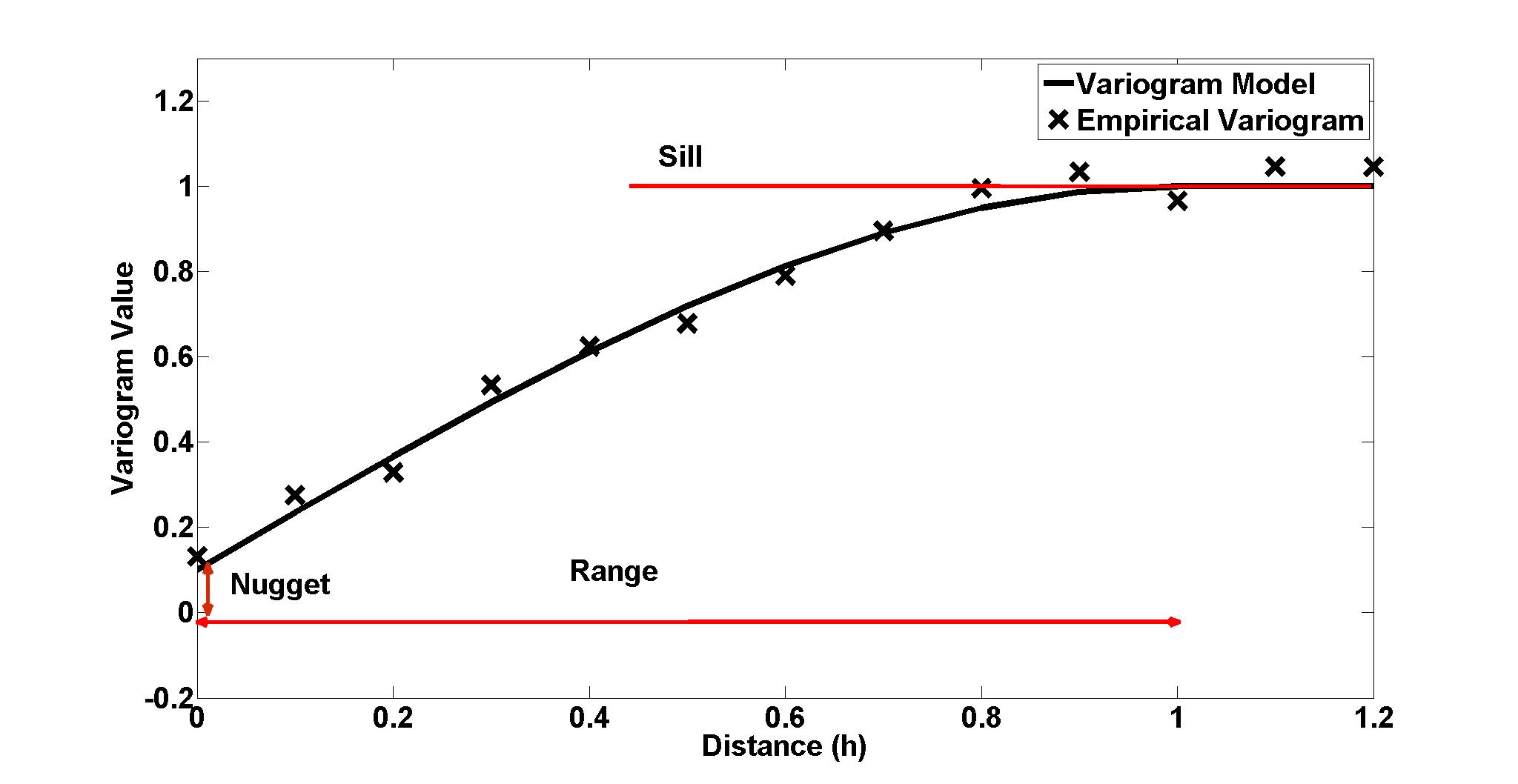}
 \caption{Example of a variogram.}
 \label{fig:variogram}
\end{figure}

\begin{itemize}
 \item Exponential Variogram 
 \begin{equation*}
  \gamma(h) = (s - n) \Big(1 - \exp\big(-\frac{h}{(ra)}\big)\Big) + n\mathbf{1}_{(0, \infty)}(h)
 \end{equation*}
 \item Spherical Variogram
 \begin{equation*}
  \gamma(h) = (s - n) \Big((\frac{3h}{2r} - \frac{h^3}{2r^3}) \mathbf{1}_{(0, r)}(h) + \mathbf{1}_{[r, \infty)}(h)\Big) + n\mathbf{1}_{(0, \infty)}(h)
 \end{equation*}
 \item Gaussian Variogram
 \begin{equation*}
  \gamma(h) = (s - n) \Big(1 - \exp\big(-\frac{h^2}{r^2a}\big)\Big) + n\mathbf{1}_{(0, \infty)}(h)
 \end{equation*}
\end{itemize}

In Exponential and Gaussian variogram models, $a$ is a free parameter that is chosen so that the variogram 
better fits the data.

\subsection{Inverse Distance Weighting}\label{sec:idw}
Inverse Distance Weighting (\ac{IDW}) is a deterministic method that estimates the value at an 
unknown position $x$ as a weighted average of values at known positions $x_1, \ldots, x_n$. 
Assuming the delta values $Z(x_1), Z(x_2), \ldots, Z(x_n)$ of representative \acp{VA} 
$x_1, x_2, \ldots, x_n$, we can estimate the delta value $Z(x)$ of a \ac{VA} at a point $x$ as  

\begin{equation}\label{eq:idw}
 \hat{Z}(x) = \Bigg\{ 
 \begin{matrix}
  \frac{\sum_{i = 1}^n w_i(x) Z(x_i)}{\sum_{i = 1}^n w_i(x)} & \forall i: D(x, x_i) \neq 0 \\
  Z(x_i) & \exists i: D(x, x_i) = 0 
 \end{matrix}
\end{equation}
where $w_i(x) = D(x, x_i)^{-p}$, and $D(\cdot, \cdot)$ is a distance function \citep{Shepard68}. 
The parameter $p$ in $w_i(x)$ is a positive real number called the ``power parameter''. The choice 
of power parameter depends on the distribution of samples and the maximum distance over 
which an individual sample is allowed to influence 
the surrounding points. Greater values of $p$ assign greater influence to values closest to the 
interpolating point. The choice of the power parameter also influences the smoothness of the 
interpolator by changing the influence radius of sample points. 

In comparison to Kriging, \ac{IDW} requires only $\Theta(n)$ 
operations to estimate the delta value 
of each new \ac{VA} contract using the delta values of $n$ representative contracts. Assuming a 
portfolio of $N$ \ac{VA} contracts, we can estimate the delta value of the portfolio by summing 
the estimated delta value of contracts in time proportional to $n \times N$. 
Hence, we expect \ac{IDW} to be faster 
than Kriging to estimate the delta value of the portfolio. The difference in speed 
is more apparent if we want a more granular view of the portfolio. In other words, if 
we are interested in the estimated delta value of each \ac{VA} contract in the portfolio, Kriging 
is much slower than \ac{IDW}. We provide further insights into this matter in Section \ref{sec:ne}. 

\subsection{Radial Basis Functions}\label{sec:rbf}
In the \ac{RBF} method, we approximate the delta value of a \ac{VA} contract $x$ as a 
weighted linear combination of radial functions centered at representative 
contracts $x_1, x_2, \ldots, x_n$:

\begin{equation}
 \hat{Z}(x) = \sum_{i = 1}^n w_i \Phi(||x - x_i||)
\end{equation} 
where $||\cdot||$ is a norm, usually chosen to be Euclidean distance.

In \ac{RBF} interpolation, the weights are chosen so that \ac{RBF} is exact at the $x_i, 1\leq i\leq n$,
points. In other words, given the values $Z(x_1), \ldots, Z(x_n)$ at points $x_1, \ldots, x_n$, the 
following linear set of equations is solved for $w_i$:

\begin{equation}\label{eq:rbf_weights}
 \begin{bmatrix}
  \Phi(||x_1 - x_1||) & \ldots & \Phi(||x_1 - x_n||) \\ 
  \vdots & \ddots & \vdots \\
  \Phi(||x_n - x_1||) & \ldots &\Phi(||x_n - x_n||)
 \end{bmatrix}
 \begin{bmatrix}
  w_1 \\ \vdots \\ w_n
 \end{bmatrix}
 = 
 \begin{bmatrix}
  Z(x_1) \\ \vdots \\ Z(x_n)
 \end{bmatrix}
\end{equation}

In our research, we chose the following commonly used radial basis functions

\begin{itemize}
 \item Gaussian
 \begin{equation}
  \Phi(x) = \exp(-\epsilon x^2)
 \end{equation}
 
 \item Multi-Quadratic 
 \begin{equation}
  \Phi(x) = \sqrt{1 + (\epsilon x)^2}
 \end{equation}

\end{itemize}

These two functions represent two classes of radial basis functions: 1) the class 
in which the value of the radial function increases with the distance from its center, 
2) the class in which the value of radial function decreases with the distance form 
its center. Although the latter class of \ac{RBF} functions, which is represented by Gaussian 
radial function in our study, seems more suitable for 
our application of interest, for the sake of completeness, we chose to experiment 
with the former class as well in our study. In both of the above-mentioned functions, 
$\epsilon$ is a free parameter that 
defines the significance of known points on the value of their neighbor 
points. 

Similar to \ac{IDW}, RBF interpolation has a running time that is proportional to $n$ for 
the delta value estimation of each \ac{VA} contract, and 
a running time of $\Theta(n \times N)$ to estimate the delta value of a portfolio 
of $N$ \ac{VA} contracts. But in addition
we need extra $\Theta(n^3)$ time to solve \eqref{eq:rbf_weights}. Hence, in total, 
the computational complexity of \ac{RBF} interpolation to estimate the delta value of a portfolio 
is $\Theta(n \times N + n^3)$. Similar to \ac{IDW}, the \ac{RBF} interpolation can provide us more granularity 
in a faster time than the Kriging method.

\section{Numerical Experiments}\label{sec:ne}
In this section, we present numerical results on the performance of 
each interpolation method in the context of the proposed framework. In all of 
our experiments, our goal is to estimate the delta value of a synthetic 
portfolio of $100,000$ \ac{VA} contracts which are chosen uniformly at random 
from the space defined by attributes listed in Table \ref{tb:portfolio}. The range of attributes 
are similar to the ones reported in \citep{Gan15, Gan13-2} which allows us to 
fairly compare our results with the reported findings in \citep{Gan15, Gan13-2}. 
However, for the sake of generality, we allow \ac{VA} contracts, with guarantee values 
that are not equal to the account value. Moreover, for \ac{VA} contracts with a GMWB rider, 
we set the guaranteed death benefit value to be equal to the guaranteed withdrawal benefit.

\begin{table}[!bt]
 \centering
 \begin{tabular}{|l|l|}
  \hline
  {\bf Attribute} & {\bf Value}\\ 
  \hline
  Guarantee Type & \{GMDB, GMDB + GMWB\} \\
  \hline
  Gender & \{Male, Female\}\\
  \hline
  Age & $\{20, 21, \ldots, 60\}$\\
  \hline
  Account Value & $[1e4, 5e5]$ \\ 
  \hline 
  Guarantee Value & $[0.5e4, 6e5]$ \\
  \hline
  Withdrawal Rate & $\{0.04, 0.05, 0.06, 0.07, 0.08\}$\\
  \hline
  Maturity & $\{10, 11, \ldots, 25\}$\\
  \hline
 \end{tabular}
 \caption{GMDB and GMWB attributes and their respective ranges of values.}
 \label{tb:portfolio}
\end{table}

In our experiments, we use the framework of \citep{Bauer08} to value each \ac{VA} contract, and 
assume the output of a \ac{MC} simulation with $10,000$ inner loop scenarios as the actual value 
of the contract. Inner loop scenarios are generated assuming a simple log-normal distribution model \citep{Hull06} 
with a risk free rate of return of $\mu = 3\%$, and volatility of $\sigma = 20\%$. Our mortality rates 
follow the 1996 I AM mortality tables provided by the Society of Actuaries.  

In the training (calibration) stage of our proposed framework, we use \ac{MC} simulations with $10,000$ inner loop 
scenarios to find the delta value of our representative contracts. The reason behind our choice is that, 
when fewer inner loop scenarios are used, e.g. $1000$ as used in \citep{Gan13-2}, we observed a noticeable 
difference between the computed delta value from successive runs. The observed difference when $1000$ is 
used can be as big as $5\%$.

\subsection{Performance}\label{sec:performance}
In this set of experiments, our objective is to provide a 
fair comparison of accuracy and computational efficiency 
of each proposed estimation method when the k-prototype distance 
function of \citep{Gan13-2} is used. Since we allow the guaranteed value of 
\acp{VA} in the synthetic portfolio to be different than their 
account value, we have changed the distance function to the following. 

\begin{equation}\label{eq:k_dist}
 D(\mathbf{x}, \mathbf{y}, \gamma) = \sqrt{\sum_{h \in N}(\frac{{x_h - y_h}}{\max_h - \min_h})^2
 + \gamma \sum_{h \in C}\delta(x_{h}, y_{h})}
\end{equation}
where $N = \{\text{AV, GD, GW, maturity, age, withdrawal rate}\}$ is the set of numerical values and 
$C = \{ \text{gender, rider} \}$ is the set of categorical values.

Similar to \citep{Gan15, Gan13-2}, we choose $\gamma = 1$. Moreover, 
we form the set of representative contracts, via the sampling method of 
Section \ref{sec:sampling_method}, from all combinations of end points 
presented in Table \ref{tb:rep_contracts1}. Because 
of the constraints on the guaranteed values, some of the entries are 
duplicate, which we remove to obtain a sample of size $1800$.

\begin{table}[!bt]
 \centering
 \begin{tabular}{|l|l|}
  \hline
  & Experiment 1\\
  \hline
  Guarantee Type & \{GMDB, GMDB + GMWB\}\\ 
  \hline
  Gender & \{Male, Female\}\\
  \hline
  Age & $\{20, 30, 40, 50, 60\}$\\
  \hline 
  Account Value & $\{1e4, 1.25e5, 2.5e5, 3.75e5, 5e5\}$\\
  \hline
  Guarantee Value & $\{0.5e4, 3e5, 6e5\}$\\
  \hline 
  Withdrawal Rate & $\{0.04, 0.08\}$\\ 
  \hline 
  Maturity & $\{10, 15, 20, 25\}$\\
  \hline
 \end{tabular}
 \caption{Attribute values from which representative contracts are generated for experiments.}
 \label{tb:rep_contracts1}
\end{table}

In order to be thorough in our experiments and comprehensive in our analysis, we present the 
results for all variants of Kriging, \ac{IDW}, and \ac{RBF} methods. For Kriging, we choose to experiment 
with all three major variogram models, i.e., spherical, exponential, and gaussian. For \ac{IDW}, we choose 
to experiment with different choices of the power parameter to see the effect of this free parameter on the 
accuracy of results. For \ac{RBF}, we study two of the most popular radial functions, Gaussian and 
multi-quadratic, and for each type of radial function, we experimented with two choices for the free 
parameter $\epsilon$. In Table \ref{tb:rel_err1}, the relative error in estimation of 
the delta value of the portfolio is presented. The relative error for method $m$ is calculated as follows.

\begin{equation}
 \text{Err}_m = \frac{\Delta_m - \Delta_{MC}}{|\Delta_{MC}|}
\end{equation}
where $\Delta_{MC}$ is the estimated delta value of the portfolio computed by \ac{MC} simulations and $\Delta_m$ is the 
estimate delta value of the portfolio computed by method $m$. While two of the Kriging methods provide accurate estimates, 
the accuracy of \ac{IDW}, and multi-quadratic \ac{RBF} methods is moderate. One interesting observation is that 
the choice of variogram model has substantial impact on the accuracy of the Kriging method and it confirms the result of 
\citep{Gan13-2} that the spherical method provides the best accuracy. Another interesting observation is the effect of 
the free parameters $p$ and $\epsilon$ on the accuracy of the \ac{IDW} and \ac{RBF} methods. The results suggest that 
the effective use of either method requires a careful tuning of these free parameters.

\begin{table}[!bt]
 \centering
 \begin{tabular}{|l|l|}
  \hline
  Method & Relative Error (\%)\\
  \hline
  Kriging (Spherical) & $-0.03$\\ 
  \hline
  Kriging (Exponential) & $-1.61$\\
  \hline 
  Kriging (Guassian) & $< -500$\\
  \hline
  IDW (p = 1) & $9.11$\\
  \hline 
  IDW (p = 10) & $13.12$\\
  \hline
  IDW (p = 100) & $11.99$\\ 
  \hline 
  RBF (Gaussian, $\epsilon = 1$) & $-1.79$\\
  \hline
  RBF (Gaussian, $\epsilon = 10$) & $37.89$\\
  \hline
  RBF(Multi-Quad, $\epsilon = 1$) & $-71.62$\\
  \hline
  RBF(Multi-Quad, $\epsilon = 10$) & $-10.86$\\
  \hline
 \end{tabular}
 \caption{Relative error in estimation of delta value via each method.}
 \label{tb:rel_err1}
\end{table}

Table \ref{tb:sim_time1} presents the running time of each algorithm in two scenarios: 1) estimating the delta value of the portfolio 
only 2) estimating the delta value of each \ac{VA} policy in the portfolio and summing them to get the delta value of 
the portfolio. While the former does not provide a granular view of the portfolio, the latter gives a more 
refined estimation process and allows for deeper analysis and insights. Note that the times 
in Table \ref{tb:sim_time1} represent only the time that it took to estimate the values once we knew the delta values 
of the representative contracts. To get the total simulation time, add $187$ seconds to these times, 
which is the time that it took to estimate the delta value of representative contracts via \ac{MC} simulations.
The results show the superiority of the proposed framework over \ac{MC} simulation (speed up $> 15\times$) 
except when Kriging is used for per policy estimation of delta. 
Because the \ac{IDW} and \ac{RBF} methods by definition require the estimation of the delta of each policy and sum the estimations to 
get the delta value of the portfolio, we can see that simulation times for these methods are approximately equal in the two presented 
scenarios. Moreover, these methods are more efficient than the Kriging method, which confirms our analysis in Section \ref{sec:sif}.

\begin{table}[!bt]
 \centering
 \begin{tabular}{|l|l|l|}
  \hline
  Method & Portfolio & Per Policy\\
  \hline
  MC & $10617$ & $10617$\\
  \hline
  Kriging (Spherical) & $312$ & $> 320000$\\ 
  \hline
  Kriging (Exponential) & $333$ & $> 320000$ \\
  \hline 
  Kriging (Guassian) & $383$ & $> 320000$\\
  \hline
  IDW (P = 1) & $285$ & $286$\\
  \hline 
  IDW (P = 10) & $288$ & $287$\\
  \hline
  IDW (P = 100) & $287$ & $301$\\ 
  \hline 
  RBF (Gaussian, $\epsilon = 1$) & $295$ & $306$\\
  \hline
  RBF (Gaussian, $\epsilon = 10$) & $294$ & $315$\\
  \hline
  RBF(Multi-Quad, $\epsilon = 1$) & $289$ & $289$\\
  \hline
  RBF(Multi-Quad, $\epsilon = 10$) & $297$ & $292$\\
  \hline
 \end{tabular}
 \caption{simulation time of each method to estimate the delta value. All times are in seconds.}
 \label{tb:sim_time1}
\end{table}

\subsection{Accuracy}\label{sec:accuracy}
The accuracy results of Table \ref{tb:rel_err1} may misleadingly 
suggest that the Kriging method with the Spherical variogram model is 
always capable of providing very accurate interpolations. In the 
experiments of this section, we provide results on the accuracy of 
different methods that contradicts this hypothesis. 

For our experiments in this section, we replicated the experiments 
of Section \ref{sec:performance}  with sets of representative contracts 
that are produced from the set of representative contracts in Section \ref{sec:performance} 
by removing $100$, $200$, $400$, $600$ and $800$ \ac{VA} contracts 
at random. Table \ref{tb:rel_err4} presents the mean and the standard deviation 
of the relative error, in estimation of the delta value of the 
\ac{VA} portfolio, for each method in these experiments. 
The results of Table \ref{tb:rel_err4} show high variance values for the  
accuracy of the Kriging methods, 
which contradicts our hypothesis. Another interesting observation 
is that the IDW methods and the RBF 
methods with a Gaussian kernel, in comparison to the Kriging methods, 
have a lower variance value for the relative error.

\begin{table}[!bt]
 \centering
 \begin{tabular}{|l|l|l|}
  \hline
  Method & Mean (\%) & STD (\%)\\ 
  \hline
  Kriging (Spherical) & $0.47$ & $1.76$\\ 
  \hline
  Kriging (Exponential) & $-0.58$ & $2.19$\\
  \hline 
  Kriging (Guassian) & $1109.02$ & $3289.81$\\
  \hline
  IDW (p = 1) & $9.14$ & $1.75$\\
  \hline 
  IDW (p = 10) & $13.14$ & $0.42$\\
  \hline
  IDW (p = 100) & $12.06$ & $0.23$\\ 
  \hline 
  RBF (Gaussian, $\epsilon = 1$) & $-1.78$ & $0.48$\\
  \hline
  RBF (Gaussian, $\epsilon = 10$) & $38.87$ & $1.42$\\
  \hline
  RBF(Multi-Quad, $\epsilon = 1$) & $-58.65$ & $16.84$\\
  \hline
  RBF(Multi-Quad, $\epsilon = 10$) & $-9.15$ & $3.56$\\
  \hline
 \end{tabular}
 \caption{Mean and standard deviation of relative error in estimation of delta value via each method.}
 \label{tb:rel_err4}
\end{table}

\subsection{Distance Function}\label{sec:df}
A key element in the definition of each estimation method is 
the choice of a distance function. While the \ac{RBF} method requires 
the choice of a proper distance function, Kriging and \ac{IDW} 
can work with any choice of distance function. A proper distance function 
satisfies non-negativity, identity of indiscernible, symmetry and 
the triangle inequality \citep{Stein09}. We call any function that has the non-negativity and a 
subset of other aforementioned properties a distance function. In this set of 
experiments, we investigate the importance of the choice of 
distance function on the accuracy of estimation for 
each interpolation method. 

To achieve this goal, we conduct two sets of experiments. In the first 
set of experiments, we study the effect of 
the $\gamma$ variable in \eqref{eq:k_dist} by reducing the value of 
$\gamma$ from $1$ to $0.05$. $\gamma$ determines the relative importance of the categorical 
attributes compared to the numerical attributes, which has not been studied previously. 
In the second set of experiments, we use the following distance function
in our method with $\gamma = 1$. 

\begin{align}\label{eq:r_dist}
&D(\mathbf{x}, \mathbf{y}, \gamma) = \sqrt{f(x_{\text{age}}, y_{\text{age}})g_{\text{age}}(\mathbf{x}, \mathbf{y}) 
+ \sum_{h \in N} g_{\text{h}}(\mathbf{x}, \mathbf{y}) + \gamma \sum_{h \in C}\delta(x_{h}, y_{h})} \nonumber\\
&f(x_{\text{age}}, y_{\text{age}}) = \exp\big(\frac{x_{\text{age}} + y_{\text{age}}}{2} - M\big) \nonumber \\
&g_h(\mathbf{x}, \mathbf{y}) = (\exp(-r_x) x_{h} - \exp(-r_y) y_{h})^2
\end{align}
where $C = \{\text{gender, rider}\}$, $N = \{\text{maturity, withdrawal rate}\}$, $r = \frac{AV}{GD}$ and $M$ is the maximum age in the portfolio. 

If we view the embedded guarantees 
in the \ac{VA} contracts as options that a policyholder can choose to exercise, 
the ratio $r$ represents the moneyness of that option. If $r \gg 1$, then 
the account value is enough to cover the amount of guaranteed value. However, if 
$r \ll 1$, the account value is insufficient to cover the guaranteed value and 
the insurer has a potential liability. Hence, in estimating the delta value for a \ac{VA} 
contract with $r \gg 1$, the delta value is close to zero and the choice of  
representative contract(s) should not affect the outcome of the estimation as long 
as the selected representative contract(s) have $r \gg 1$. The choice of function 
$g(\cdot,\cdot)$ in \eqref{eq:r_dist} 
captures the aforementioned idea. In addition, as the age of the policyholder increases 
their mortality rate also increases (consult the data of 1996 I AM 
mortality table). Hence, the liability and delta value of similar contracts which 
differ only in the age of the policyholder increases with age. Because of this, more 
emphasis should be placed on estimating the delta value for senior policyholders, 
which is the motivation behind the introduction of the function $f(\cdot,\cdot)$ in \eqref{eq:r_dist}.

\begin{table}[!bt]
 \centering
 \begin{tabular}{|l|l|l|}
  \hline
  & \multicolumn{2}{|c|}{Relative Error (\%)}\\\cline{2-3}
  Method & Experiment 1 & Experiment2\\
  \hline
  Kriging (Spherical) & $1.94$ & $*$\\ 
  \hline
  Kriging (Exponential) & $-0.37$ & $*$\\
  \hline 
  Kriging (Guassian) & $< -500$ & $*$\\
  \hline
  IDW (p = 1) & $8.97$ & $-4.87$\\
  \hline 
  IDW (p = 10) & $13.21$ & $3.90$\\
  \hline
  IDW (p = 100) & $11.99$ & $2.32$\\ 
  \hline 
  RBF (Gaussian, $\epsilon = 1$) & $-2.56$ & $*$\\
  \hline
  RBF (Gaussian, $\epsilon = 10$) & $37.89$ & $*$\\
  \hline
  RBF(Multi-Quad, $\epsilon = 1$) & $-35.74$ & $*$\\
  \hline
  RBF(Multi-Quad, $\epsilon = 10$) & $-6.88$ & $*$\\
  \hline
 \end{tabular}
 \caption{Relative error in the estimation of the delta value by each method. In experiment 1, \eqref{eq:k_dist} is used with $\gamma = 0.05$, and 
 in experiment 2, \eqref{eq:r_dist} is used with $\gamma = 1$. ``$*$'' indicates that the method cannot work with the choice of distance 
 function because it causes singularities in the computations.}
 \label{tb:rel_err2}
\end{table}

Table \ref{tb:rel_err2} presents the accuracy of our estimation by each method in both experiments. 
In experiment one, the choice of $\gamma = 0.05$, in general, has improved the accuracy of most interpolation 
schemes. Kriging interpolation with a Spherical variogram, the \ac{IDW} method with $P = 10$, and the \ac{RBF} 
method with Gaussian kernel and $\epsilon = 1$ are the only 
schemes for which the accuracy deteriorated. In experiment two, the Kriging and \ac{RBF} methods  
encounter singularities with \eqref{eq:r_dist}; however, the choice of \eqref{eq:r_dist} has improved 
the accuracy of the \ac{IDW} methods. In general, it seems that the choice of distance function and free parameters
plays a key role in the accuracy of the interpolation schemes.

\subsection{Variogram}\label{sec:vario}
As mentioned in Section \ref{sec:kriging}, Kriging methods 
work with variogram models. The choice of variogram model 
is dictated by its closeness to the empirical variogram. In the 
previous experiments, we showed that we can have better results using 
a  spherical variogram model; however, we haven't provided any analysis 
supporting why this variogram is a better choice. 
In this section, we address this subject. In particular, we conduct 
experiments to explore whether we can increase the accuracy of 
the Kriging method by choosing a variogram function that can 
better approximate the empirical variogram.

To compute the empirical variogram, we partition the x-axis into 
20 intervals of equal length $\frac{h_{\max}}{20}$ where $h_{\max}$ is the 
maximum distance between two \ac{VA} policies using the distance function \eqref{eq:k_dist} 
and with $\gamma = 1$. 
In each interval, to approximate \eqref{eq:empirical_vario}, 
we use the average of the squared difference 
of the delta value of all pairs of \ac{VA} policies that have a distance that falls 
into that interval as the representative for the empirical variogram for that interval. 
We call the piece-wise linear function that is formed by connecting the representative
value for each interval the empirical variogram.

\begin{figure}[!bt]
 \centering
 \subfigure[Spherical variogram.]{ 
  \includegraphics[trim=0cm 1cm 6cm 0cm, clip=true, width=0.22\textwidth, height=0.2\textheight]{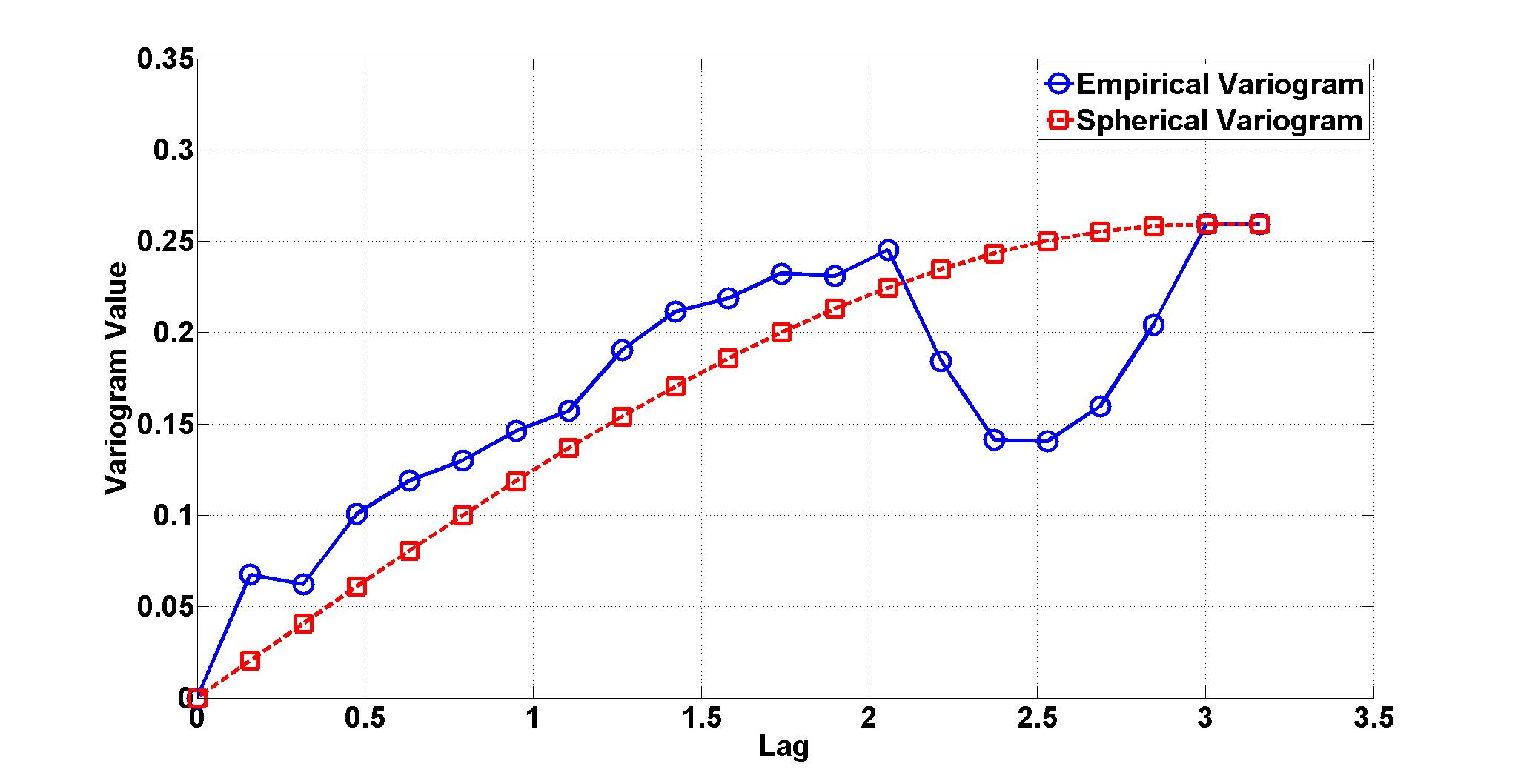}  
  \label{fig:sphVario} 
 }
 \subfigure[Exponential variogram.]{ 
  \includegraphics[trim=0cm 1cm 6cm 0cm, clip=true, width=0.22\textwidth, height=0.2\textheight]{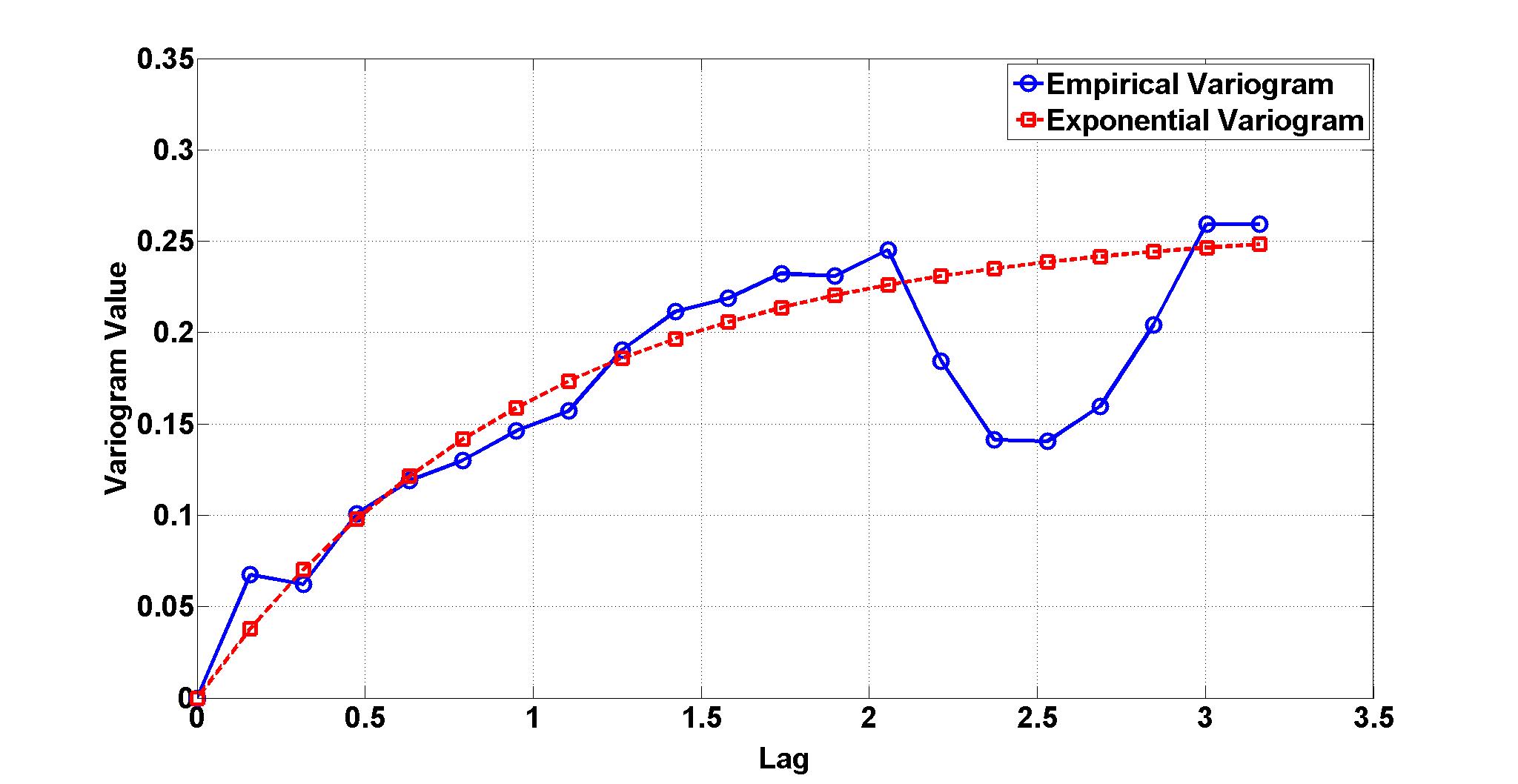}  
  \label{fig:expVario} 
 }
  \subfigure[Gaussian variogram.]{ 
  \includegraphics[trim=0cm 1cm 6cm 0cm, clip=true, width=0.22\textwidth, height=0.2\textheight]{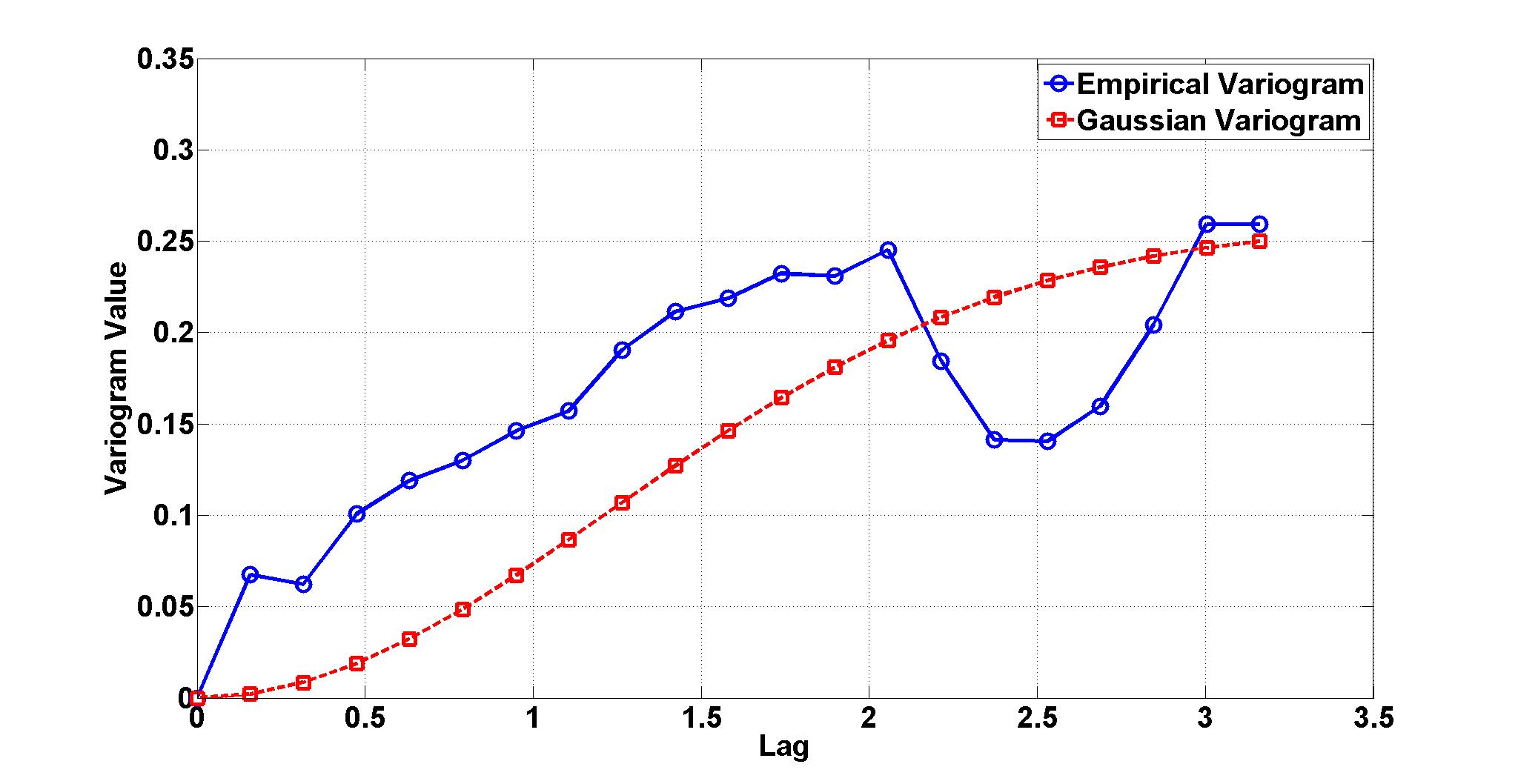}  
  \label{fig:gauVario} 
 }
 \subfigure[Linear approximation.]{ 
  \includegraphics[trim=0cm 1cm 6cm 0cm, clip=true, width=0.22\textwidth, height=0.2\textheight]{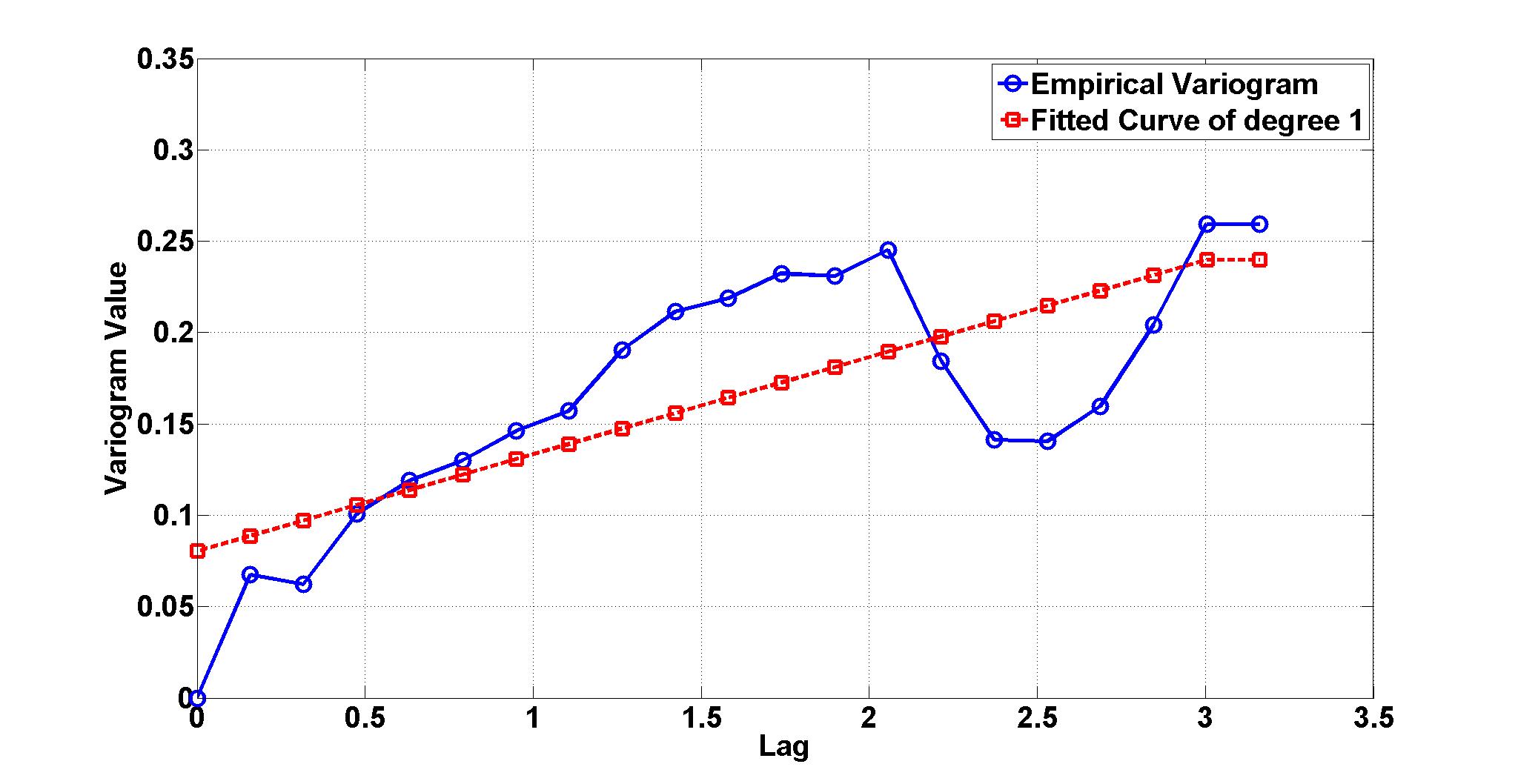}  
  \label{fig:deg1Vario} 
 }
 \subfigure[Quadratic approximation.]{ 
  \includegraphics[trim=0cm 1cm 6cm 0cm, clip=true, width=0.22\textwidth, height=0.2\textheight]{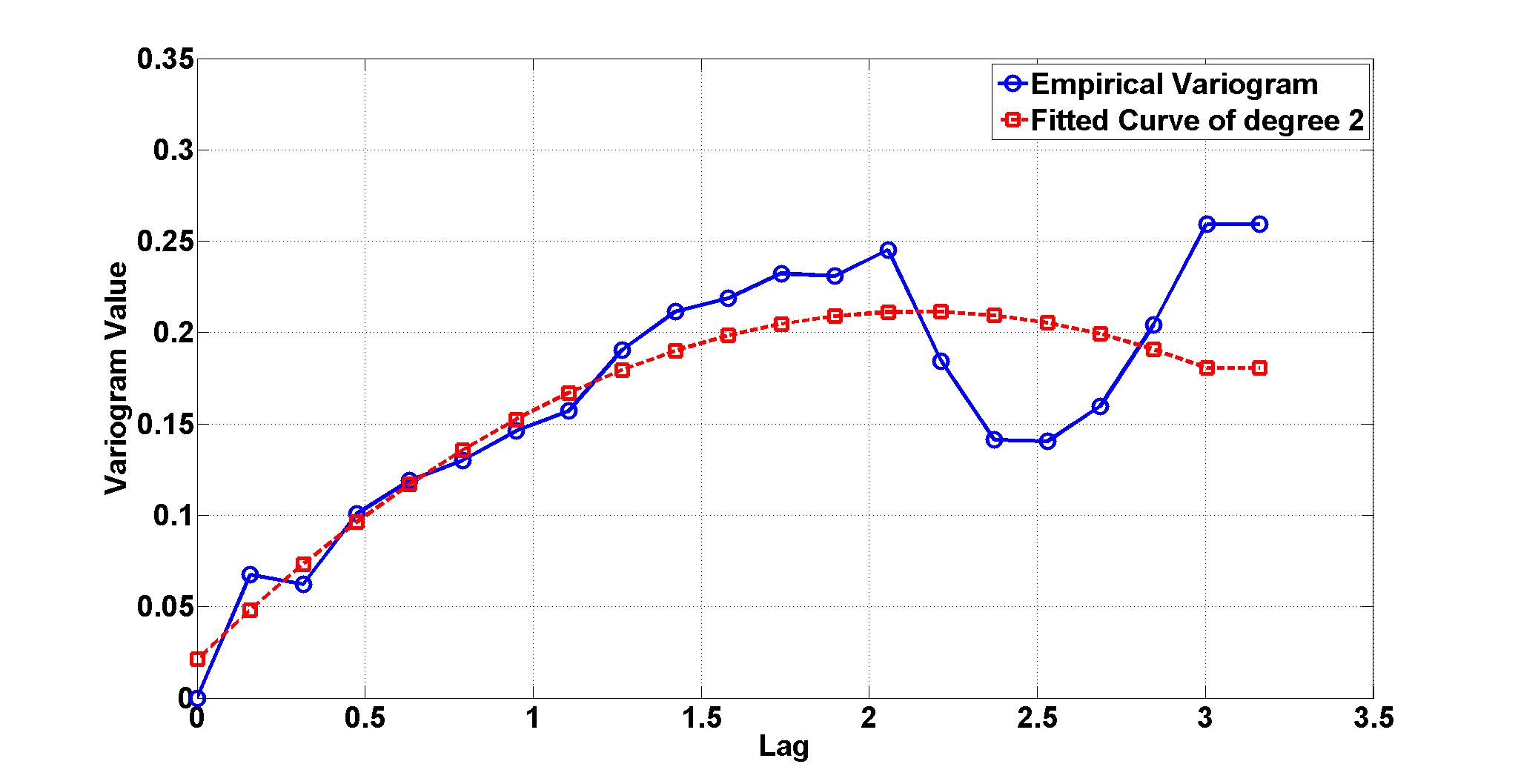}  
  \label{fig:deg2Vario} 
 }
 \subfigure[Cubic approximation.]{ 
  \includegraphics[trim=0cm 1cm 6cm 0cm, clip=true, width=0.22\textwidth, height=0.2\textheight]{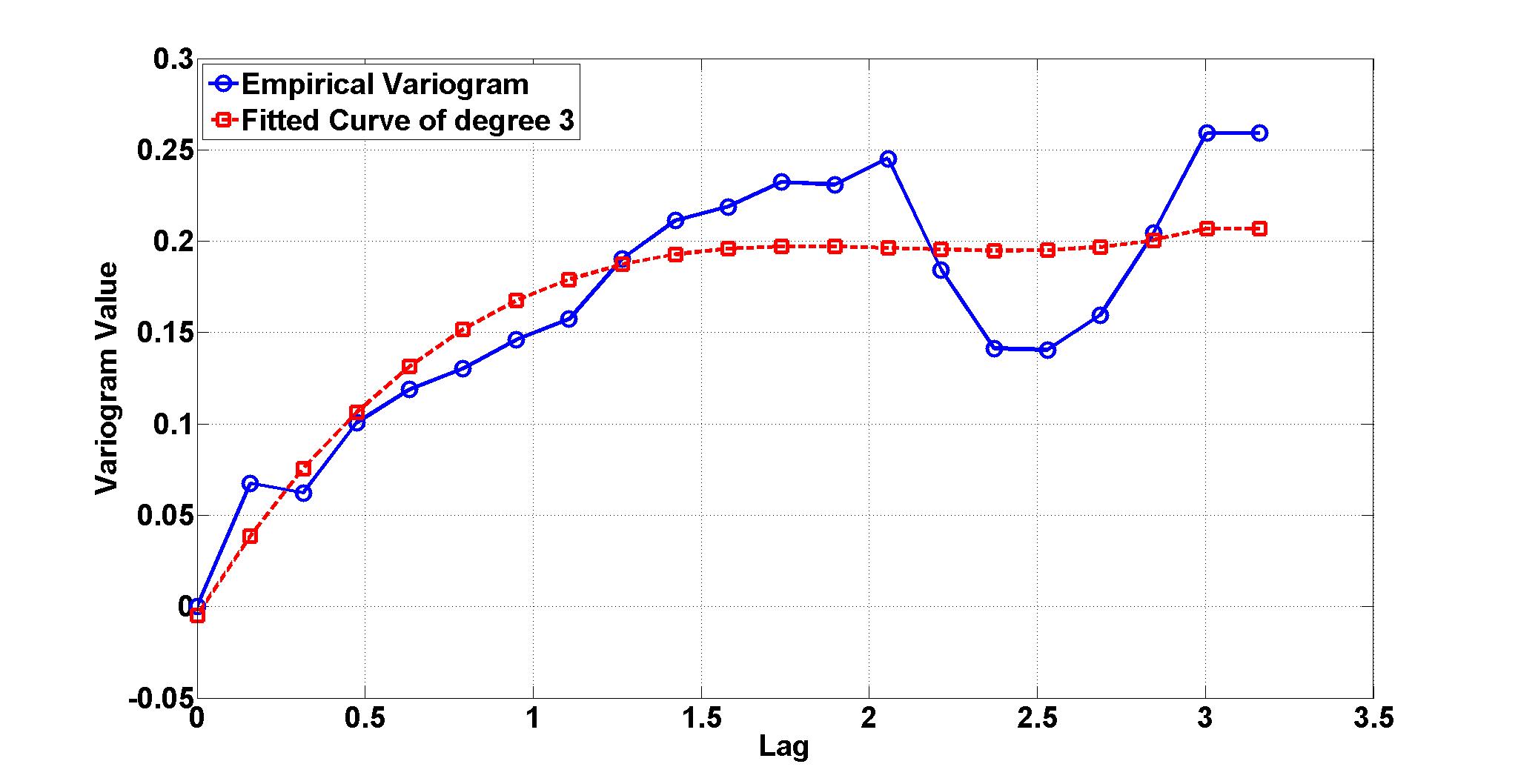}  
  \label{fig:deg3Vario} 
 }
 \subfigure[Quartic approximation.]{ 
  \includegraphics[trim=0cm 1cm 6cm 0cm, clip=true, width=0.22\textwidth, height=0.2\textheight]{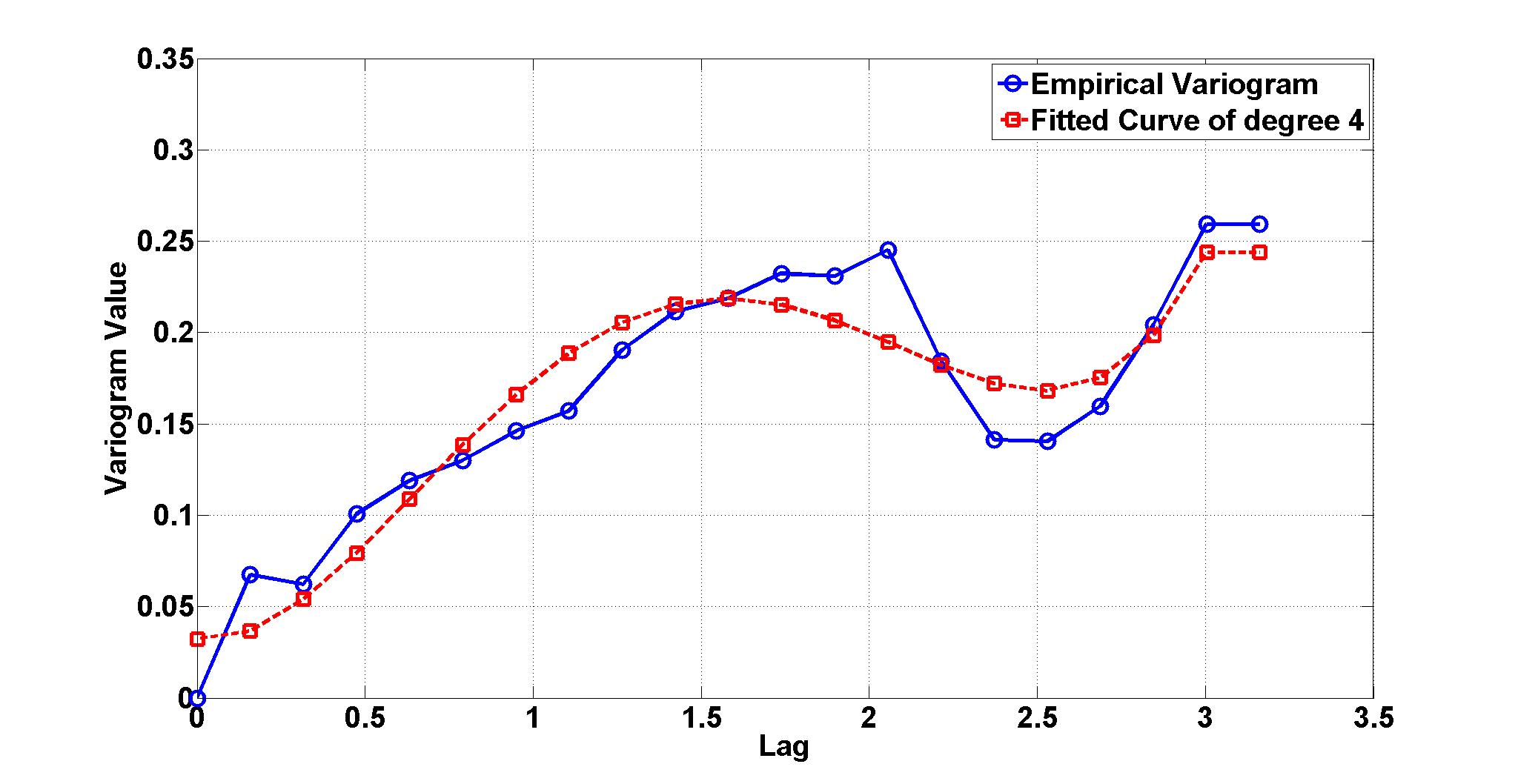}  
  \label{fig:deg4Vario} 
 }
 \caption{Comparing the variogram models with the empirical variogram.}
 \label{fig:vario2}
\end{figure}

To approximate the empirical variogram, we use polynomials of degree $1$, $2$, $3$ 
and $4$. The polynomials are best MSE approximations of the empirical variogram in 
the interval between zero and range. At any distance above the range, the estimated 
value is assumed to be the value of the polynomial at the range (Figure \ref{fig:vario2}). 
This is necessary in order to have a proper variogram model \citep{Cressie93} 
without producing jumps in the variogram.  

The accuracy of Kriging using each approximation of the empirical variogram is presented in Table \ref{tb:rel_err3}. 
An interesting, yet counter intuitive, observation is that the accuracy of Kriging is worst for the 
quartic MSE approximation variogram model that best fits the empirical variogram. Even comparing the graph of 
the exponential and spherical variogram models with the empirical variogram, the exponential variogram model seems 
to fit the empirical variogram model better than the spherical variogram model, but the accuracy of Kriging with 
the exponential variogram model is worse than the accuracy of Kriging with the spherical variogram model.

\begin{table}[!bt]
 \centering
 \begin{tabular}{|l|l|}
  \hline
  Method & Relative Error (\%)\\
  \hline
  Spherical & $-0.03$\\ 
  \hline
  Exponential & $-1.61$\\
  \hline 
  Guassian & $< -500$\\
  \hline
  Deg 1 & $-6.00$\\
  \hline 
  Deg 2 & $-32.17$\\
  \hline
  Deg 3 & $-2.78$\\
  \hline
  Deg 4 & $< -500$\\
  \hline
 \end{tabular}
 \caption{Relative error in estimation of the delta value via Kriging with different variogram models.}
 \label{tb:rel_err3}
\end{table}

\begin{figure}[!bt]
 \centering
 \includegraphics[trim=0cm 1cm 4.5cm 0cm, clip=true, width=0.7\textwidth, height=0.3\textheight]{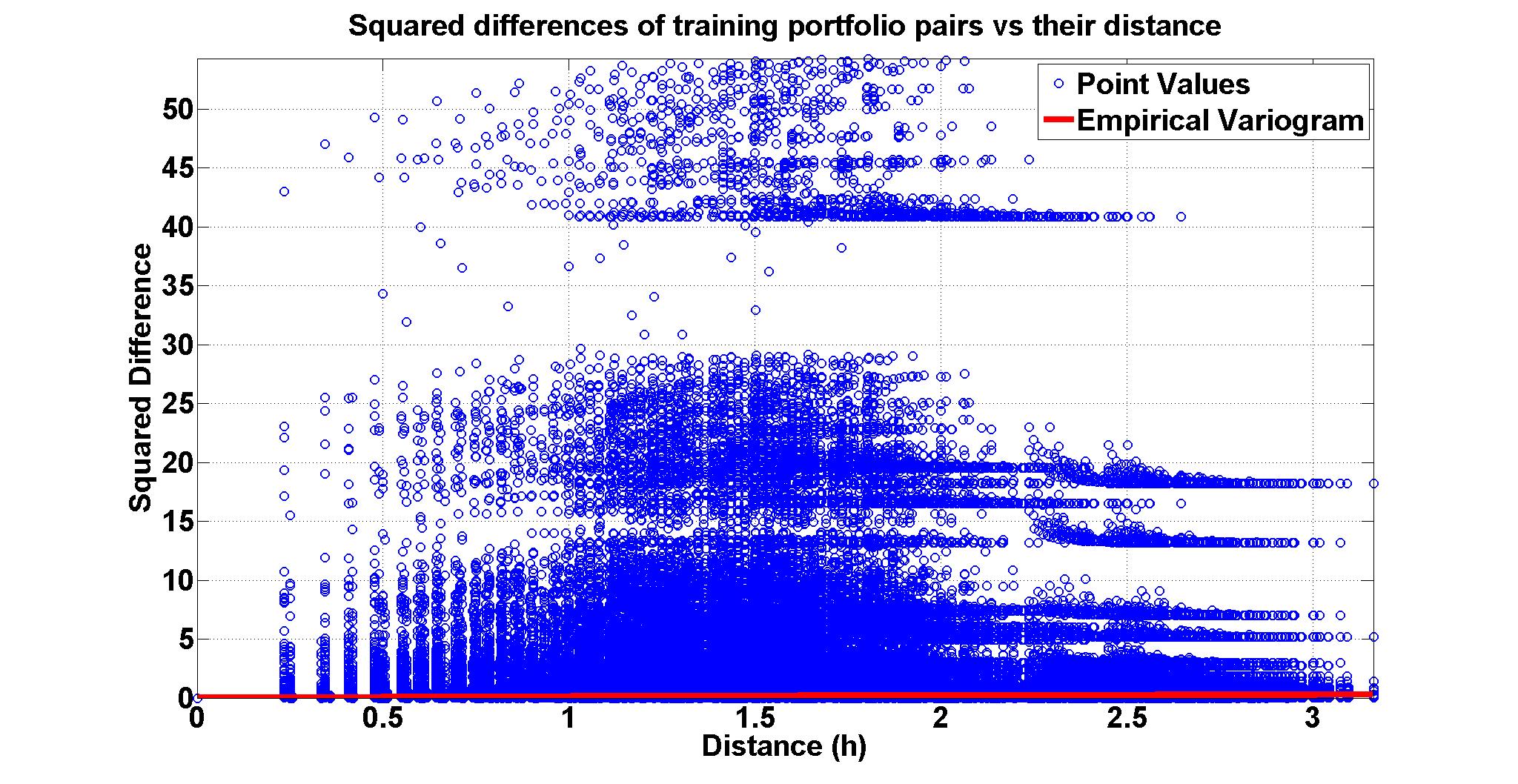}
 \caption{Squared difference of delta values of \ac{VA} pairs in representative contracts.}
 \label{fig:squared_diff}
\end{figure}

Because of these counter intuitive results, we took a closer look at the data from which the 
empirical variogram was generated. Figure \ref{fig:squared_diff} shows a graph of the squared differences of delta 
values of a pair of \ac{VA} contracts versus their distance from each other. Surprisingly, the point values 
do not look similar to their average, i.e., the empirical variogram. We expected to see a graph similar to 
Figure \ref{fig:variogram} where the point values are in close proximity to the empirical variogram and the variogram model. 
However, the data do not suggest the existence of any pattern from which a variogram model can be estimated. 
In particular, the data contradict the second-order 
stationary assumption underlying the Kriging method, and hence brings into question the appropriateness of the Kriging 
method for our application of interest.

\section{Concluding Remarks}\label{sec:conclusion}
As we discussed in Sections \ref{sec:intro} and \ref{sec:ne}, valuing a large 
portfolio of \ac{VA} contracts via nested \ac{MC} simulations 
is a time consuming process \citep{Fox13, Reynolds08}. Recently, 
Gan and Lin \citep{Gan13-2} proposed a Kriging framework that ameliorates 
the computational demands of the valuation process. 
The proposed method is a special case of a more general framework 
called spatial interpolation. The spatial interpolation framework works 
as follows: A group of representative \ac{VA} contracts is selected via a 
sampling method. For each contract in the sample, the Greeks are estimated 
via a \ac{MC} simulation. Then a spatial interpolation method is used to find the 
Greeks for each contract in the portfolio as a linear combination of the Greeks 
of the contracts in the sample. 

Two integral parts of the spatial interpolation framework are 
the choice of sampling method and the choice of interpolation scheme. In this paper, 
we studied various interpolation methods (i.e., Kriging, \ac{IDW}, \ac{RBF}) 
that can be used in this framework and 
postponed the discussion of the choice of sampling method to a future paper. 
In our experiments, we focused on a synthetic portfolio of \ac{VA} contracts with 
GMDB and GMWB riders and compared the efficiency and accuracy of interpolation methods 
to estimate the delta value of this portfolio. As we discuss in detail in Section 
\ref{sec:ne}, while Kriging can provide better accuracy than the \ac{IDW} and \ac{RBF} methods, 
it has a slower running time than either of these deterministic methods, 
because Kriging has to solve \eqref{eq:ord_krig} that in general takes $\Theta(n^3)$ time ($n$ is 
the number of samples). The computational cost of Kriging is exacerbated if we 
compute the delta value of each \ac{VA} policy in the portfolio. 
In general, a more granular view of the portfolio with Kriging comes at the expense of a 
much longer running time. 

Our experiments also provided more insights into the importance of the choice of a distance 
function in the accuracy of various interpolation schemes. While Kriging methods, with 
exponential and spherical variogram models, can provide relatively accurate estimates 
using \eqref{eq:k_dist}, they fail, due to numerical instability, to provide any estimates 
using \eqref{eq:r_dist}. However, in comparison to \eqref{eq:k_dist}, \eqref{eq:r_dist} 
allows \ac{IDW} methods to provide better estimates. Our experiments 
also showed that achieving the best result with either \eqref{eq:k_dist} or \eqref{eq:r_dist} 
requires fine tuning the free parameters in these distance functions or in the 
methods themselves. 

An interesting observation from our experiments is the non-existence of any pattern that 
supports the validity of the second-order stationary assumption that underlines the Kriging method. 
Our observation was the result of our experiments to accurately estimate the empirical variogram. 
Our hope was that a more accurate estimation of the empirical variogram would increase the accuracy of 
Kriging methods. Despite our initial belief, the 
experiments showed no relation between the accuracy of the Kriging method and the closeness of 
the variogram model to the empirical variogram. Hence, it is not clear why the Kriging method provides 
accurate estimates despite the fact that the second-order stationary assumption does not seem to hold for 
the Greeks of the \ac{VA} portfolio.

From our results, none of traditional spatial interpolation methods enjoy all three properties of 
accuracy, efficiency, and granularity. While Kriging based methods are accurate and fairly efficient at 
the portfolio level, they lack efficiency when a granular view of the portfolio is required. On the other hand, 
\ac{IDW} and \ac{RBF} based methods are efficient, and can provide a granular view of the portfolio, but 
they are not as accurate as Kriging based methods. In addition, the accuracy of all methods is dependent 
on the choice of an appropriate distance function, which requires tuning from expert users. It is 
not yet clear to us what is the best approach to find an appropriate distance function for each method. 
In a future paper, we will discuss how we can circumvent this issues and still 
achieve accuracy, efficiency, and granularity via a neural network approach to spatial interpolation. 
The proposed framework resolves the issues of the choice of distance function by 
learning an appropriate distance function to be used for each portfolio given the characteristics of 
the portfolio. 

In the future, we also plan to address the problem of choosing an effective sampling method via a novel approach that uses 
statistical characteristics of the input portfolio to provide the output sample from the space 
in which the input portfolio is defined.

\section*{Acknowledgements}
This research was supported in part by the Natural Sciences and Engineering Research Council of Canada (NSERC).

\bibliographystyle{model2-names}
\bibliography{Risk-Paper1}







\end{document}